\documentclass[iop,linenumbers]{emulateapj}
\usepackage{color} 
\usepackage{ulem}
\usepackage{amsmath}
\usepackage{enumerate}
\usepackage{apjfonts}
\usepackage{subfigure}
\usepackage{float}
\usepackage{multirow}
\usepackage[colorlinks=true,citecolor=blue,breaklinks=true]{hyperref}
\usepackage[toc,page]{appendix}
\usepackage{soul}
\setstcolor{red}
\usepackage{longtable}
\usepackage{rotating}

\usepackage{threeparttable}

\begin{document}

\title{estimating the spin of MAXI J1348--630 from intermediate and soft states using \textit{Insight}-HXMT data}

\author{J. Guan\altaffilmark{1}, R.C. Ma\altaffilmark{1,2}, L. Tao\altaffilmark{1}, A.C. Fabian\altaffilmark{3}, J.A. Tomsick\altaffilmark{4}, S.N. Zhang\altaffilmark{1,5}, L. Zhang\altaffilmark{1}, J.L. Qu\altaffilmark{1}, S. Zhang\altaffilmark{1}}

\altaffiltext{1}{Key Laboratory of Particle Astrophysics, Institute of High Energy Physics, Chinese Academy of Sciences, Beijing 100049, China}
\altaffiltext{2}{Dongguan Neutron Science Center, 1 Zhongziyuan Road, Dongguan 523808, China}
\altaffiltext{3}{Institute of Astronomy, University of Cambridge, Madingley Road, Cambridge CB3 0HA, UK}
\altaffiltext{4}{Space Sciences Laboratory, 7 Gauss Way, University of California, Berkeley CA 94720-7450, USA}
\altaffiltext{5}{University of Chinese Academy of Sciences, Chinese Academy of Sciences, Beijing 100049, China}
\email{jguan@ihep.ac.cn; maruican@ihep.ac.cn}
\shorttitle{spin of MAXI J1348--630}

\shortauthors{Guan et al.}

\begin{abstract}
We present a broadband spectral analysis of the soft-intermediate and soft states using \textit{Insight}-HXMT observations of the black hole binary MAXI J1348--630 during its outburst in 2019. We employ a combination of reflection and continuum fitting methods to measure the spin of the black hole and determine a spin of $a=0.79\pm0.13$, which is consistent with most of the previous results. Additionally, we investigate some sources of systematic uncertainty that could bias the measurement. The valid spectral hardening factor can partially account for the spin evolution observed during the outburst. Other potential factors, such as assumptions about the corona's structure and their interaction with reflected photons, can also affect the accuracy of spin measurements. Furthermore, our analysis reveals that MAXI J1348--630 exhibits a significantly high disc density ($\sim10^{22}\,\rm{cm}^{-3}$), with a moderate iron abundance of approximately 2.5 times solar.

\end{abstract}


\section{Introduction}
\label{sec:intro}
Black hole (BH) X-ray binaries (BHXRBs) are binary systems consisting of a BH that accretes matter from a companion star via the Roche-lobe overflows. These systems spend most of the time in the quiescent state, and occasionally undergo outburst phases for weeks to months with X-ray intensity several orders of magnitude stronger than that in the quiescent state \citep{Lasota2001}. During a typical outburst, the BH evolves through various spectral states \citep{2010LNP...794...53B,2006ARA&A..44...49R}, usually, the sequence from the Hard State (HS) to the Hard-Intermediate State (HIMS), then to the Soft-Intermediate State (SIMS), followed by the Soft State (SS), and then back to SIMS, HIMS, and eventually the Hard State (HS), forming a ‘q’ shaped track in the hardness-intensity diagram \citep[HID;][]{2004MNRAS.355.1105F,2005Ap&SS.300..107H}. It is believed that the transitions of the spectral states characterised by the relative strength of the thermal and non-thermal X-ray emission are driven by the changes in the accretion rate, the disc-corona geometry or a combination of all these factors \citep{2013ApJ...763...48R}. Thus the BH spectral states in turn provide insights on the accretion flow geometry such as the disc inner radius, the coronal region size and the evolution of them \citep{2019MNRAS.484.1972J}. The standard paradigm for the HS is a truncated disc and an inner advective flow \citep{1997ApJ...489..865E}. The spectrum is featured by a Comptonisation (power-law) component and a strong reflection component, and sometimes a weak thermal component. In the intermediate states, whether the disc is truncated is a debated topic. Some evidence show that the inner accretion disc shrinks \citep{2011MNRAS.415.2323I,2014MNRAS.442.1767P} while others support a collapsing corona with the inner disc reaching the BH’s innermost stable circular orbit (ISCO) \citep{2015ApJ...813...84G,2019Natur.565..198K}. The X-ray spectrum here is a soft power-law tail and reflection emission coexisting with a strong thermal component. It is well established that the inner disc extends to the ISCO in the SS state, which is the basic assumption of most spin measurements \citep{2010ApJ...718L.117S}. The X-ray spectrum is dominated by a strong thermal component emitted from the inner accretion disc.

Spin is one of the two parameters describing an astrophysical BH. It changes the geometry of space-time and the ways a BH interacting with the surrounding environment, thus is important to constrain the BH physics \citep{2007arXiv0707.4492M}. Since the spin influences the gravitational well of the BH which changes the X-ray spectral shape and luminosity of the disc emission \citep{2006ApJ...652..518M} and distorts the reflection radiation \citep{2016AN....337..362D}, estimations of the BH spin could be achieved either by modeling the thermal emission of the disc \citep[the continuum-fitting (CF) method;][]{1997ApJ...482L.155Z, McClintock2011} or the reflection component \citep[the reflection spectroscopy;][]{1989MNRAS.238..729F,2003PhR...377..389R}. Both methods fundamentally assume that the inner disc extends to the ISCO. Since $R_{\rm ISCO}/(GM/c^2)$ simply and monotonically maps the spin parameter $a_{*}$ ($a_{*}=cJ/GM^{2}$, where $J$ is the BH angular momentum and $M$ is the BH mass), by estimating the inner disc radius $R_{\rm in}$, so long as it is equivalent to $R_{\rm ISCO}$, one could directly derive the BH spin $a_{*}$.

An important issue of spin measurement is the error budget. Several systematic uncertainties can bias the measurement and might lead to conflicting results obtained from the two spin measurement methods. The main systematic uncertainties in the CF method are the uncertainties of the BH mass $M$, the distance $D$, and the inclination of the accretion disc. High accuracy of these parameters is necessary for this method to succeed. Large uncertainties or wrong estimation of these parameters would lead to poor or wrong constrain of the spin. Additionally, \cite{2021MNRAS.500.3640S} employed an indirect method to study how uncertainties in the spectral hardening factor \( f_{\text{col}} \) impact spin measurements. This method reverse-engineered a published black hole spin marginal density to obtain the disc flux normalization marginal density and then compared it with that calculated from different combinations of model parameters, including the black hole mass, distance, inner disc inclination, spectral hardening factor \( f_{\text{col}} \) and the inner disc radius. The authors demonstrated that the uncertainties in the spectral hardening factor \( f_{\text{col}} \)  can introduce modest systematic errors into spin measurements due to the complexities inherent in the accretion disc. However, this approach allows one to explore how choosing different hardening factor \( f_{\text{col}} \) affects the spin error budget but cannot decide which one is preferred by the observed data. Thus the value of \( f_{\text{col}} \) adopted in the actual disc continuum fitting still appeals to a theoretically motivated value, e.g. in the modern continuum fitting technique {\tt KERRBB2} \citep{McClintock2014}, one would first tabulate \( f_{\text{col}} \) on an ($l$, $a$)-grid by fitting the color-corrected {\tt KERRBB} model \citep{2005ApJS..157..335L} to the sophisticated  {\tt bhspec} disc atmosphere model \citep{2005ApJ...621..372D} and then select a most appropriate one for each {\tt KERRBB} fit to the observed data. The systematic uncertainties in the reflection method can come from the spectral models. The disc photosphere is treated as a constant density slab in the radiative transfer calculations. Although it remains unclear to what extent it affects the spin measurement, this oversimplification has significantly affected the estimation of other parameters in some situations. Statistically acceptable solutions are obtained with high density disc and solar-like iron abundance or low density disc and supper high iron abundance \citep{2018ApJ...855....3T}. Scattering and absorption by more distant matter or line-of-sight gas can mimic the signatures of the disc reflection. Correcting for this effect to accurately derive the inner disc reflection from observed spectra presents some challenges, especially when the X-ray corona source is partially obscured \citep{2021ARA&A..59..117R}.




MAXI J1348--630 (hereafter ‘MAXI J1348’) is a bright X-ray transient discovered by MAXI/GSC on January 26, 2019 \citep{2019ATel12425....1Y,2020ApJ...899L..20T}. Its radio and optical counterparts are identified later \citep{2019ATel12456....1R,2019ATel12430....1D}. It exhibits first a giant outburst, which traces a typical ‘q’ shape track in the HID, and several subsequent hard state re-brightening acting like ‘failed outbursts’ \citep{2020ApJ...899L..20T,2021MNRAS.505.3823Z}. All types of low-frequency QPOs as well as the rise, quenching and re-activation of compact jets are detected through the different spectral phases of the outburst \citep{2020MNRAS.496.4366B,2021MNRAS.505.3823Z,2021MNRAS.505..713J,2021MNRAS.504..444C,2021MNRAS.505L..58C,2022MNRAS.511.4826C}.
MAXI J1348’s multi-wavelength properties strongly suggest that the source is a BHXRB \citep[][and references therein]{2021MNRAS.505.3823Z}. The geometrical distance of MAXI J1348 is estimated to be $3.39\pm0.34$\,kpc based on a study of a giant X-ray dust scattering ring. This leads to a revised mass estimation for the BH of $M=11\pm2\,M_{\odot}$ \citep{2021A&A...647A...7L}. \cite{2022MNRAS.511.4826C} obtains a jet inclination angle of ${29.3^{+2.7}_{-3.2}}^{\circ}$ by modeling the jet motion with a dynamical external shock model. A broadband spectral analysis of \textit{NuSTAR} data with high density disc reflection models results in close-to solar iron abundance and an inclination of $\sim30^{\circ}$\citep{2021MNRAS.508..475C}. There have been six prior estimates of MAXI J1348's spin from X-ray observations. The first of these, $a=0.78^{+0.04}_{-0.04}$, is derived from a joint fitting of the SIMS and SS \textit{NuSTAR} data using relxill-based reflection model \citep{2022MNRAS.511.3125J}. \cite{2022MNRAS.513.4869K} obtained a consistent result of  $a=0.80\pm{0.02}$ from the SS using \textit{NICER} and \textit{NuSTAR} observations. \cite{2023MNRAS.526.6041S} also derived a moderate high spin of $a=0.82^{+0.04}_{-0.03}$ using \textit{Insight}-HXMT IMS and SS data via the reflection method. However, the \textit{AstroSat} data supports a nearly maximal spin \citep{2023RAA....23a5015M}. \cite{2024ApJ...969...40D} also presented evidence of an extremely rapidly spinning black hole, based on a relxill-based reflection analysis of all \textit{NuSTAR} data. In contrast to these findings, a moderate spin value of $a=0.42^{+0.13}_{-0.50}$, featuring a notably broader uncertainty range, was constrained from the very soft state in \textit{Insight}-HXMT data, employing the CF method \citep{2023MNRAS.522.4323W}.


In this paper, in order to understand the causes behind these conflicting results, we perform detailed spectral analysis of MAXI J1348 with \textit{Insight}-HXMT and one set of simultaneous \textit{NuSTAR} data to investigate the spin and accretion flow of the system. The SIMS and SS are preferred since the inner disc has reached the ISCO during the state \citep{2022ApJ...927..210Z}. In order to characterize both the soft thermal emission and the disc reflection self-consistently, we combine the CF method and reflection spectroscopy method to constrain the spin. 


The paper is organized as follows. Section~\ref{sec:obs} describes the observations and data reduction of \textit{Insight}-HXMT and \textit{NuSTAR}. Section~\ref{sec:results} provides the details of the spectral analysis. We present the discussion and conclusion in Section~\ref{sec:dis} and ~\ref{sec:con}.

\section{Observations and data reduction}
\label{sec:obs}
\subsection{\textit{Insight}-HXMT data reduction}
\label{sec:insight data reduction}

\textit{Insight}-HXMT is the China's first X-ray astronomy satellite, successfully launched on 2017 June 15 \citep{2020SCPMA..6349502Z}. There are three payloads on board: the low energy X-ray telescope \citep[LE, $1-15$\,keV, 384\,$\rm{cm}^{2}$,][]{2020SCPMA..6349505C}, the medium energy X-ray telescope \citep[ME, $5-35$\,keV, 952\,$\rm{cm}^{2}$,][]{2020SCPMA..6349504C}, and the high energy X-ray telescope \citep[HE, $20-250$\,keV, 5100\,$\rm{cm}^{2}$,][]{2020SCPMA..6349503L}. \textit{Insight}-HXMT monitored the entire outburst of MAXI J1348 from 2019 January to July with a total exposure time of 1500\,ks. 

In this paper, we perform the spectral analysis of MAXI J1348 during its SIMS and SS (from MJD 58522 to MJD 58587) with 31 \textit{Insight}-HXMT observations, since the inner radius of the disc has reached the ISCO \citep{2022ApJ...927..210Z}, and the count rates of ME and HE are not too small and follow a smooth evolution trend. The detailed information for these observations are listed in Table \ref{tab:obs_inf}. The data are extracted using the \textit{Insight}-HXMT Data Analysis software ({\sc hxmtdas}) v2.04 with the following recommended criteria: (1) the offset for the pointing position is $\le0.05^{\circ}$; (2) the geomagnetic cutoff rigidity is $>8$\,GV; (3) the elevation angle is $>6^{\circ}$; (4) the extraction time is at least $300$\,s away from the South Atlantic Anomaly passage. Only the data from detectors with small field of views (FOVs) are used to avoid possible contamination from nearby sources and the bright Earth. The backgrounds for the three payloads are estimated using the blind detectors because the spectral shapes of the particle backgrounds are the same for both the blind and small FOV detectors \citep{2020JHEAp..27...24L,2020JHEAp..27...44G,2020JHEAp..27...14L}. The energy bands adopted for spectral analysis are 1--1.6, 1.9--10\,keV for LE, 10--35\,keV for ME and 35--150\,keV for HE. We exclude energy between $1.6-1.9$\,keV of LE to avoid sharp instrumental features. For LE, the spectra are re-binned with at least 100 counts per bin while the ME spectra before and after $21$\,keV, every 2 and 5 channels are rebinned into one bin, respectively. For HE, every 2, 4 and 34 channels are re-binned into one bin before, during and after $130.6-188.4$\,keV, respectively.

\subsection{\textit{NuSTAR} data reduction}
\label{sec:nustar data reduction}

\textit{NuSTAR} \citep{Harrison2013} possess two identical detectors, referred to as focal plane module FPMA and FPMB, and can cover the energy band from 3 to 79 keV. In this work, we use a \textit{NuSTAR} observation of MAXI J1348--630 with the ObsID 80402315008 (MJD 58525.3), which is simultaneous with the ObsID P021400201510$\sim$1513 from \textit{Insight}-HXMT. 

We processed the \textit{NuSTAR} data using the \textit{NuSTAR} data analysis software ({\sc nustardas}) v2.0.0 with CALDB version v20210305. We used \textit{nupipeline} to calibrate and clean the event files. Due to the source being too bright, with a count rate greater than 100\,cts\,{$\rm s^{-1}$}, we set \texttt{statusexpr=``STATUS==b0000xxx00xxxx000''} during this process. When extracting the source spectrum with \textit{nuproducts}, we selected a circular region with a radius of 250 arcsec, centering the source within it. The background region was selected to be as far away from the source as possible, with a circular radius of 100 arcsec. The response files were also generated using the \textit{nuproducts} command. We used the \textit{grppha} command to ensure that each bin contains a minimum of 50 photons.

\begin{table*}
    \caption{\textit{Insight}-HXMT Observations of MAXI J1348 during the SIMS and SS.}
    \label{tab:obs_inf}
    \begin{center}
    \begin{tabular}{ccccccccc}
\hline
ObsID & Observed date & Observed date & LE exposure &  LE rate & ME exposure & ME rate  &  HE exposure  & HE rate \\
      & (MJD) &   & (s) & ($\rm {cts~s^{-1}}$) & (s)  &  ($\rm {cts~s^{-1}}$)  & (s) & ($\rm {cts~s^{-1}}$) \\
\hline
1101$\sim$1103$^{\rm{a}}$ & 58522.59 & 2019-02-08T14:10:48 & 1773 & $3565.2\pm{1.4}$ & 5571 & $200.8\pm{0.2}$ & 6200  & $221.9\pm{0.8}$ \\
1501$\sim$1505$^{\rm{b}}$ & 58523.79 & 2019-02-09T18:50:08 & 3427 & $3514.8\pm{1.0}$ & 9097 & $210.8\pm{0.2}$ & 10676 & $236.1\pm{0.6}$ \\
1506$\sim$1509 & 58524.77 & 2019-02-10T18:31:05 & 2417 & $3389.6\pm{1.2}$ & 5578 & $166.0\pm{0.2}$ & 4732  & $181.9\pm{0.9}$ \\
1510$\sim$1513 & 58525.32 & 2019-02-11T07:40:07 & 2030 & $3365.7\pm{1.3}$ & 5357 & $165.2\pm{0.2}$ & 7537  & $180.7\pm{0.7}$ \\
1601$\sim$1603 & 58526.17 & 2019-02-12T04:07:01 & 1394 & $3250.2\pm{1.5}$ & 3142 & $131.7\pm{0.2}$ & 3422  & $138.9\pm{1.1}$ \\
1604$\sim$1607 & 58526.60 & 2019-02-12T14:23:32 & 2512 & $3206.3\pm{1.1}$ & 3853 & $150.1\pm{0.2}$ & 5572  & $178.4\pm{0.8}$ \\
1701$\sim$1704 & 58527.30 & 2019-02-13T07:09:29 & 3561 & $3126.4\pm{0.9}$ & 5415 & $193.4\pm{0.2}$ & 7350  & $225.5\pm{0.8}$ \\
1801$\sim$1803 & 58528.62 & 2019-02-14T14:58:00 & 3679 & $3047.2\pm{0.9}$ & 5326 & $115.3\pm{0.2}$ & 5153  & $152.0\pm{0.9}$ \\
2101$\sim$2103 & 58533.13 & 2019-02-19T03:06:20 & 3703 & $2761.8\pm{0.9}$ & 4472 & $113.1\pm{0.2}$ & 2490  & $139.5\pm{1.3}$ \\
2401 & 58538.17 & 2019-02-24T03:58:12 & 5503 & $2452.1\pm{0.7}$ & 7143 & $111.77\pm{0.15}$ & 9793  & $141.0\pm{0.7}$ \\
2510$\sim$2512 & 58540.73 & 2019-02-26T17:27:04 & 5918 & $2260.5\pm{0.6}$ & 10930 & $88.81\pm{0.11}$ & 10247  & $119.4\pm{0.7}$ \\
2601$\sim$2602 & 58542.74 & 2019-02-28T17:43:34 & 3484 & $2013.0\pm{0.8}$ & 9958 & $65.60\pm{0.11}$ & 9514  & $94.8\pm{0.7}$ \\
2701 & 58544.07 & 2019-03-02T01:32:47 & 1816 & $2082.3\pm{1.1}$ & 5236 & $89.8\pm{0.2}$ & 6953  & $120.2\pm{0.8}$ \\
2801 & 58546.72 & 2019-03-04T17:11:48 & 703 & $1913.2\pm{1.7}$ & 4305 & $78.4\pm{0.2}$ & 4514  & $97.3\pm{1.0}$ \\
2901 & 58547.91 & 2019-03-05T21:50:36 & 4086 & $1817.5\pm{0.7}$ & 8008 & $65.84\pm{0.12}$ & 9233  & $94.0\pm{0.7}$ \\
3001$\sim$3003 & 58548.84 & 2019-03-06T20:07:31 & 5343 & $1847.6\pm{0.6}$ & 9293 & $80.07\pm{0.12}$ & 9780  & $108.2\pm{0.6}$ \\
3101$\sim$3104 & 58550.30 & 2019-03-08T07:08:22 & 5561 & $1650.9\pm{0.6}$ & 6988 & $55.19\pm{0.12}$ & 7992  & $80.4\pm{0.7}$ \\
3201 & 58551.56 & 2019-03-09T13:22:39 & 3651 & $1669.2\pm{0.7}$ & 3520 & $62.8\pm{0.2}$ & 4846  & $86.3\pm{1.0}$ \\
3601 & 58553.41 & 2019-03-11T09:56:05 & 1975 & $1627.4\pm{0.9}$ & 2162 & $66.2\pm{0.2}$ & 1718  & $80.7\pm{1.7}$ \\
3801 & 58555.27 & 2019-03-13T06:29:04 & 180 & $1438\pm{3}$ & 1050 & $46.7\pm{0.3}$ & 796  & $70\pm{2}$ \\
4201 & 58559.58 & 2019-03-17T13:52:53 & 3232 & $1292.6\pm{0.6}$ & 3035 & $26.49\pm{0.15}$ & 1584  & $43.0\pm{1.6}$ \\
4301 & 58560.51 & 2019-03-18T12:08:57 & 3052 & $1285.6\pm{0.7}$ & 2969 & $50.5\pm{0.2}$ & 3347  & $76.7\pm{1.1}$ \\
4401 & 58561.43 & 2019-03-19T10:24:59 & 3232 & $1282.9\pm{0.6}$ & 2982 & $52.4\pm{0.2}$ & 4892  & $71.3\pm{1.0}$ \\
4501 & 58562.36 & 2019-03-20T08:41:01 & 2840 & $1179.5\pm{0.7}$ & 2748 & $45.9\pm{0.2}$ & 3738  & $68.1\pm{1.1}$ \\
4601 & 58565.08 & 2019-03-23T01:53:44 & 120 & $1105\pm{3}$ & 1179 & $28.4\pm{0.3}$ & 913  & $44\pm{2}$ \\
4801$\sim$4803 & 58566.94 & 2019-03-24T22:26:00 & 2274 & $1058.5\pm{0.7}$ & 4298 & $26.38\pm{0.13}$ & 4489 & $36.5\pm{1.0}$ \\
4901 & 58568.13 & 2019-03-26T03:04:01 & 718 & $1035.6\pm{1.2}$ & 1747 & $26.6\pm{0.2}$ & 645  & $42\pm{3}$ \\
5101 & 58571.84 & 2019-03-29T20:09:43 & 419 & $933.0\pm{1.5}$ & 1723 & $44.9\pm{0.2}$ & 2441  & $67.3\pm{1.3}$ \\
5501 & 58574.03 & 2019-04-01T00:40:31 & 120 & $921\pm{3}$ & 2082 & $34.4\pm{0.2}$ & 984  & $56\pm{2}$ \\
5601$\sim$5604 & 58576.68 & 2019-04-03T16:20:06 & 1676 & $848.3\pm{0.7}$ & 6818 & $18.26\pm{0.09}$ & 8277  & $28.6\pm{0.7}$ \\
5701$\sim$5703 & 58578.87 & 2019-04-05T20:51:08 & 1616 & $810.9\pm{0.7}$ & 4787 & $13.54\pm{0.11}$ & 4683  & $25.3\pm{0.9}$ \\
5801$\sim$5802 & 58580.79 & 2019-04-07T18:59:51 & 3289 & $773.8\pm{0.5}$ & 4526 & $19.57\pm{0.12}$ & 5208  & $34.7\pm{0.9}$ \\
5901$\sim$5904 & 58582.58 & 2019-04-09T13:57:10 & 5487 & $665.1\pm{0.3}$ & 7179 & $21.53\pm{0.09}$ & 8682  & $38.1\pm{0.7}$ \\
6001$\sim$6005 & 58583.58 & 2019-04-10T13:48:53 & 4517 & $706.8\pm{0.4}$ & 7137 & $13.13\pm{0.09}$ & 7524  & $26.5\pm{0.7}$ \\
6101 & 58586.96 & 2019-04-13T22:56:00 & 2120 & $653.5\pm{0.6}$ & 2782 & $18.39\pm{0.15}$ & 3532  & $33.9\pm{1.1}$ \\
\hline   
\end{tabular}
\end{center}
\footnotesize{$^{\rm{a}}$ short for P021400201101$\sim$P021400201103.}\newline
\footnotesize{$^{\rm{b}}$ except P021400201504.}\newline
\end{table*}

\section{Spectral Analysis and Results}
\label{sec:results}

The spectra are analysed with XSPEC V12.11.1. The adopted models here include the galactic absorption effect by implementing the TBABS model component with \citet{2000ApJ...542..914W} abundances and \citet{1996ApJ...465..487V} cross sections. An additional CONSTANT component is included to account for the calibration discrepancies between different payloads (throughout the paper, we fix the multiplicative factor of LE and FPMA to 1). The fitting procedure minimizes the $\chi^2$ goodness-of-fit statistic. All uncertainties calculated for the spectral parameters are at $90\%$ confidence level, unless noted particularly. Considering the accuracy of the calibration (e.g.,  the uncertainties of the instrumental responses and the background), a systematic error of $1.5\%$ is added in the \textit{Insight}-HXMT spectral fitting \citep{2020JHEAp..27...64L}, which have no effect on the fitting parameters.

\subsection{spectral properties}
\label{sec:spectral properties}

\begin{figure}
    \includegraphics[width=\columnwidth]{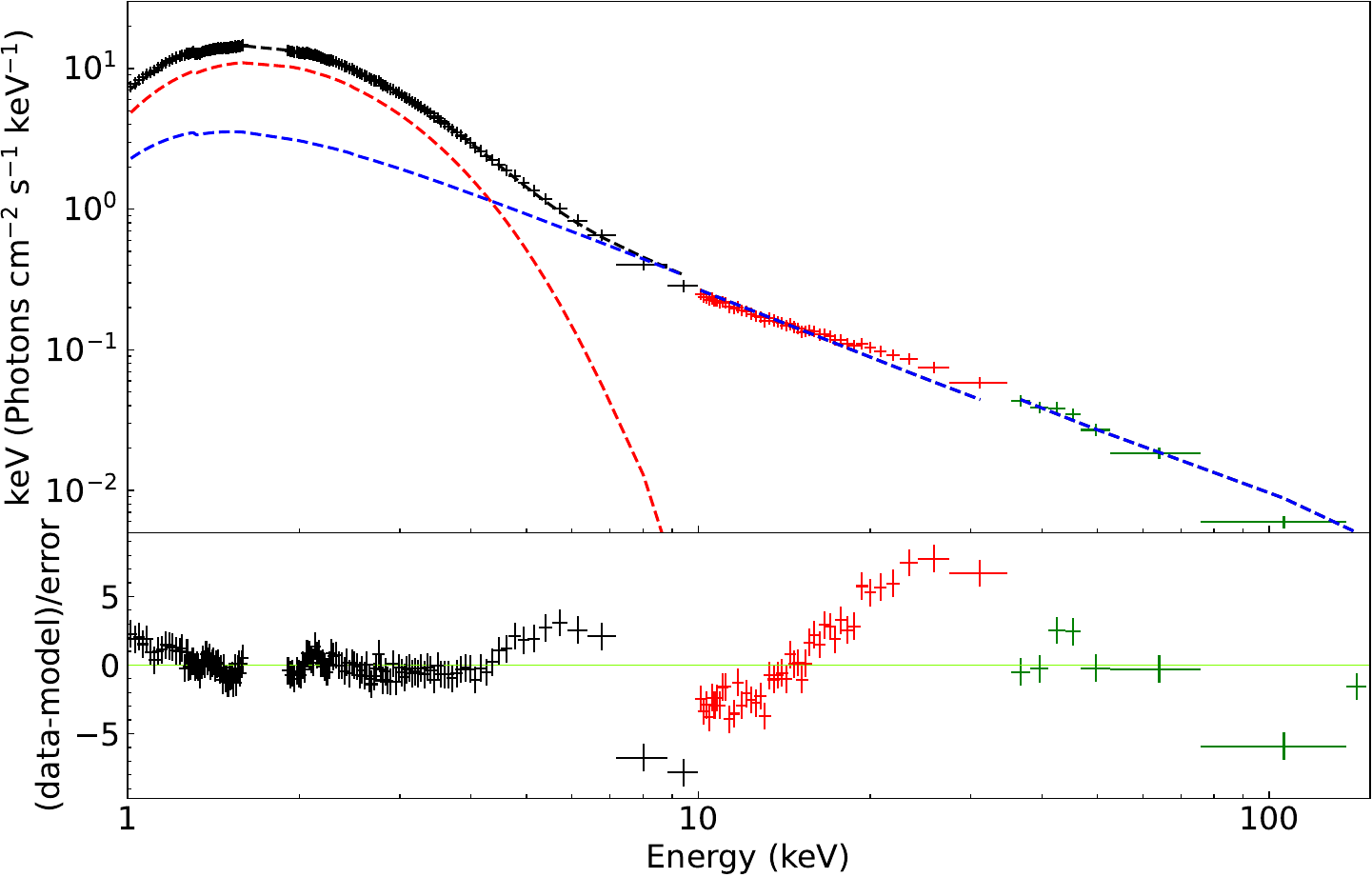}
    \caption{A representative spectrum (ObsID P021400203601) of MAXI J1348 fitted with {\tt constant*TBabs*(diskbb+powerlaw)}. The red, blue, and black dashed lines represent {\tt diskbb}, {\tt powerlaw}, and the total component, respectively.}
    \label{fig:spec_diskpl}
\end{figure}

We start our analysis by performing preliminary spectral fits to the \textit{Insight}-HXMT spectra with an absorbed multicolour disc blackbody \citep[{\tt diskbb},][]{1984PASJ...36..741M} plus a power-law component, in specific, {\tt constant*TBabs*(diskbb+powerlaw)}. It is clear that the model does not fit the data well. A broad iron line as well as an obvious Compton hump in the residues indicate the presence of relativistic reflection (Figure \ref{fig:spec_diskpl}).

Then in order to derive the spin of MAXI J1348 and account for the reflection component, we replace {\tt diskbb} with {\tt kerrbb2} \citep{2006ApJ...652..518M} to fit the disc thermal emission and replace {\tt powerlaw} with a more physical model {\tt Nthcomp} \citep{1996MNRAS.283..193Z, 1999MNRAS.309..561Z} to fit the thermally comptonized continuum, and use a {\tt Nthcomp} continuum-based high density reflection model {\tt reflionx\_hd\_Nthcomp} \citep[][John Tomsick, private communication]{2005MNRAS.358..211R, 2018ApJ...855....3T} convolved with smeared relativistic accretion disc line profiles using {\tt relconv} \citep{2011IAUS..275..100D} to fit the reflection emission. The total set-up of the model is {\tt const*Tbabs*(kerrbb2+NthComp+relconv*\\reflionx\_hd\_Nthcomp)}. 

The hybrid code {\tt kerrbb2} is a modified version of the fully relativistic thin-disc model KERRBB \citep{2005ApJS..157..335L}. It retains the special features of {\tt kerrbb} (e.g. frame dragging, Doppler boosting, gravitational redshift, returning radiation, and light bending), and incorporates the effects of spectral hardening via a pair of look-up tables for $f_{\rm col}$ corresponding to two values of the viscosity parameters: $\alpha=0.01, 0.1$ \citep{2006ApJ...652..518M}. The look-up table is computed with {\tt bhspec} \citep{2005ApJ...621..372D} based on non-LTE atmosphere models within an  $\alpha$-viscosity prescription. We adopt $\alpha=0.1$ \citep{2011MNRAS.416..941S}, which is also a typical value (0.1--0.2) as determined by numerical simulations of the magneto-rotational instability \citep[e.g.,][]{Hirose2014, Coleman2016}, throughout this work. Thus, the model {\tt kerrbb2} has just two fitting parameters: the spin $a_{*}$ and the mass accretion rate $\dot{M}$. We fix the torque at the inner boundary of the disc to zero, turn on the effect of the returning radiation, set the normalization to unity and fix the input parameters $M$, $i$ and $D$ to $11\,M_{\odot}$, $29.3^{\circ}$ and $3.39$\,kpc, respectively (for more details on parameters selection see Section~\ref{sec:intro}). The spin and inclination of {\tt relconv} are linked to that of {\tt kerrbb2}. The other parameters in {\tt relconv} are fixed at their default values. The spectral index $\Gamma$, the electron temperature $kT_{\rm e}$ and the seed photon temperature $kT_{\rm bb}$ of {\tt Nthcomp} are linked to that of {\tt reflionx\_hd\_Nthcomp}. Since $kT_{\rm e}$ can not be constrained well, we fix it to 500\,keV. The limb-darkening is turned off, and we have checked that this factor seldom affects the results. The hydrogen column density is fixed at $1 \times 10^{22}\,\rm{cm}^{-2}$ in all observations to avoid introducing possible evolution trends of the spin. The \textit{Insight}-HXMT spectra prefer a hydrogen column density of $\sim1 \times 10^{22}\,\rm{cm}^{-2}$ rather than the value $0.86 \times 10^{22}\,\rm{cm}^{-2}$ \citep{2020ApJ...899L..20T} according to $\chi^2$, however we found that it affects little on the spin measurement.

\begin{figure}
\centering
    \includegraphics[width=\columnwidth]{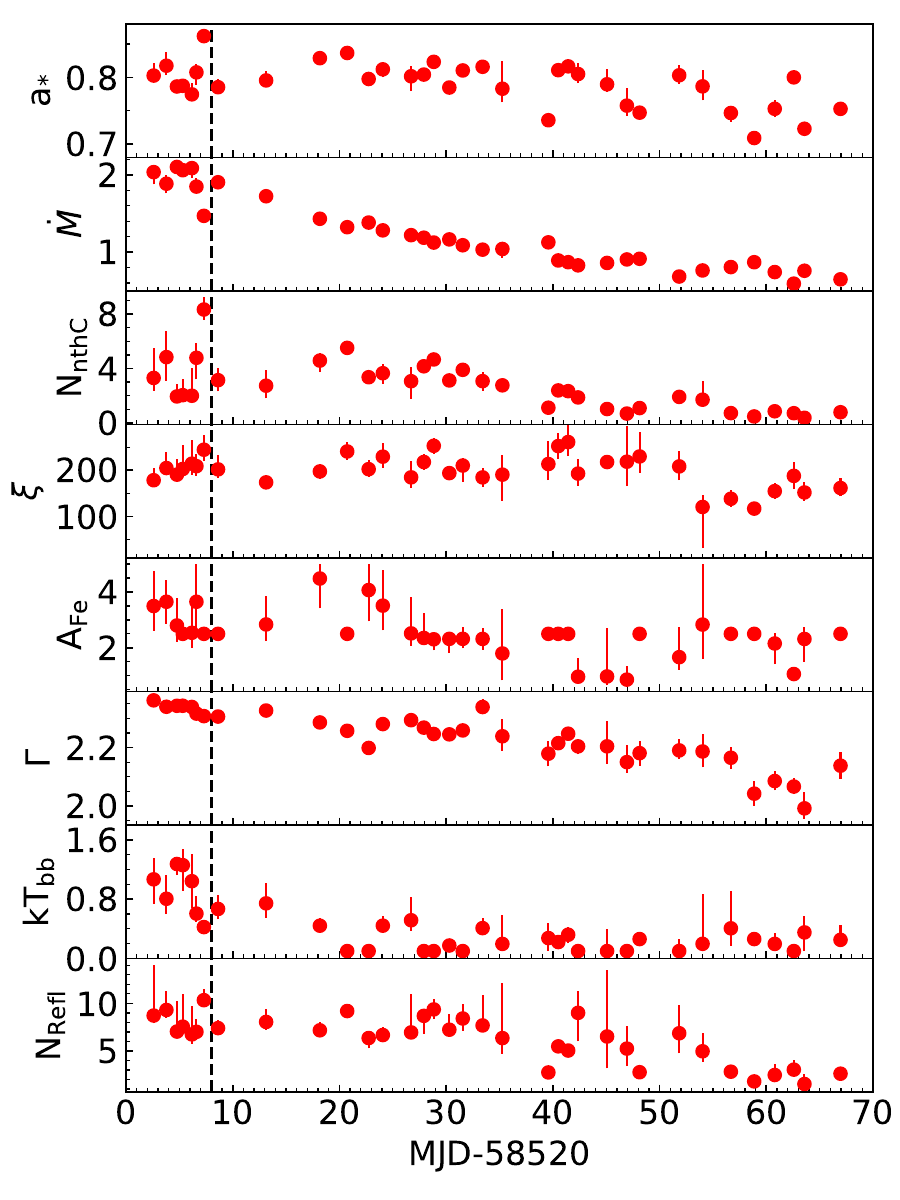}
    \caption{Evolution of the spectral parameters of model {\tt const*Tbabs*(kerrbb2+NthComp+relconv*\\reflionx\_hd\_Nthcomp)}. $a_{*}$ is the spin; $\dot{M}$ is the mass accretion rate in units of $10^{18}~\rm{g~s^{-1}}$; $N_{\rm nthC}$ is the normalization of {\tt NthComp}; $\xi$ is the ionization of the accretion disc; $A_{\rm{Fe}}$ is the iron abundance; $\Gamma$ is the photon index; $kT_{\rm{bb}}$ is the seed photon temperature; $N_{\rm Refl}$ is the normalizaton of {\tt reflionx\_hd\_Nthcomp}. The vertical line represents the transition from SIMS to SS on approximately MJD 58258.}
    \label{fig:pars}
\end{figure}

\begin{figure}
    \includegraphics[width=\columnwidth]{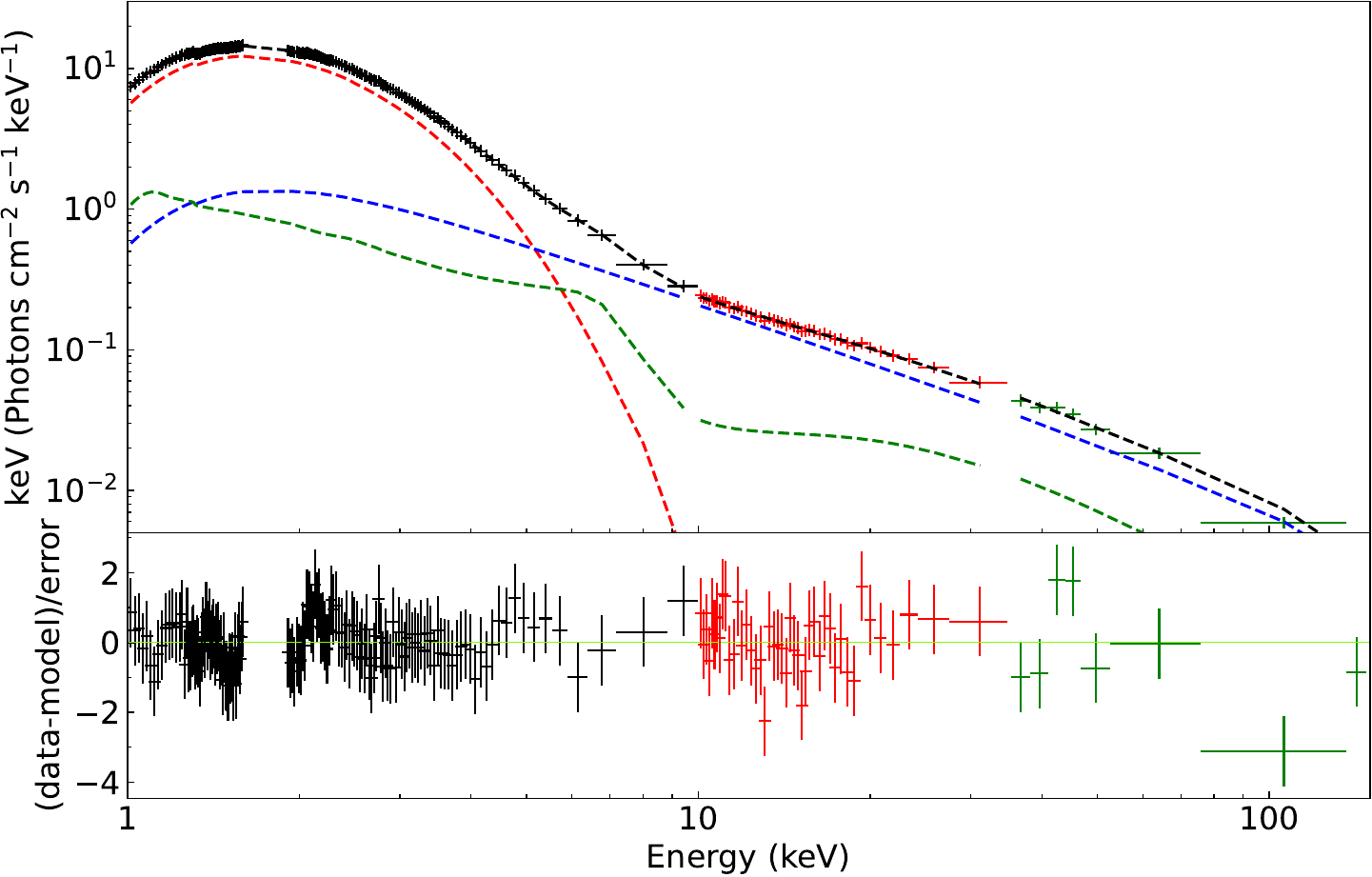}
    \caption{A representative spectrum (ObsID P021400203601) of MAXI J1348 fitted with {\tt const*Tbabs*(kerrbb2+NthComp+relconv*\\reflionx\_hd\_Nthcomp)}. The red, blue, green, and black dashed lines represent the thermal, Compton, reflection and total components, respectively.}
    \label{fig:spec_krbb}
\end{figure}

\begin{figure}
    \includegraphics[width=1.1\columnwidth]{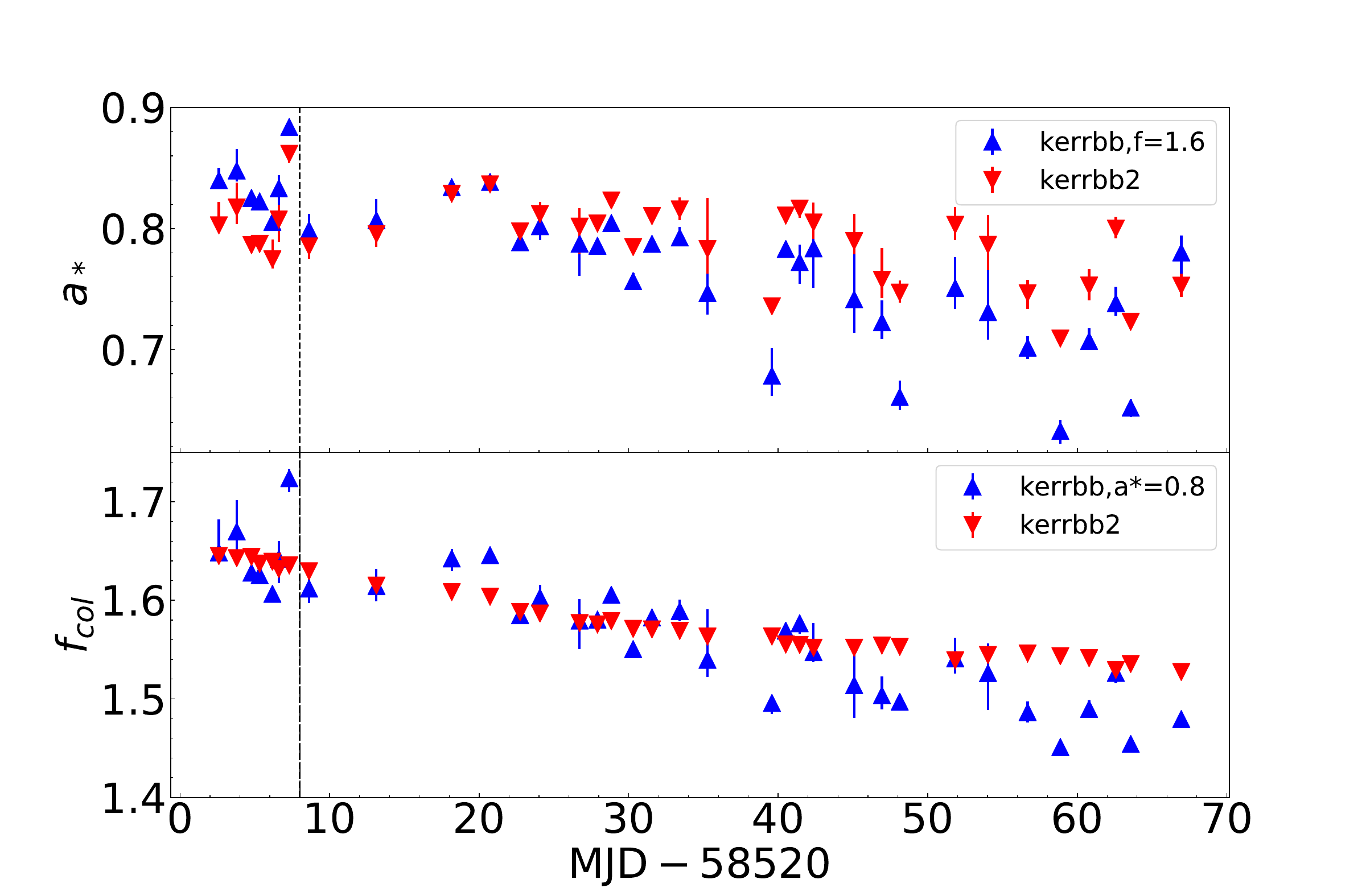}
    \caption{Top panel: Spin measurements are shown with different hardening factors $f_{\rm col}$. The blue and red represent the results of the {\tt kerrbb} model with $f_{\rm col}$ fixed at 1.6 and the {\tt kerrbb2} model with a free $f_{\rm col}$, respectively. Bottom panel: The evolution of the hardening factor $f_{\rm col}$ over time is depicted. The blue and red represent the results of the {\tt kerrbb} model with a fixed spin at 0.8 and the {\tt kerrbb2} model with a freely fitted spin, respectively.}
   \label{fig:compare_f}
   \end{figure}

The fitting results are shown in Figure~\ref{fig:pars} and Table~\ref{tab:fit_pars}. Reasonable fits are obtained with the reduced $\chi^2$ ranging from $0.62$ to $1.02$. A representative plot of the fitting spectrum is given in Figure~\ref{fig:spec_krbb} with no obvious residuals. The obtained CONSTANT components of ME and HE are $\sim1$. Except that the mass accretion rate shows a decreasing evolution trend, other fitting parameters are quite consistent across time. The deduced disc luminosity calculated with $L=\eta\dot{M}c^{2}$  also decreases, with an Eddington-scaled luminosity $l=L/L_{\rm Edd}$ ranging from 0.162 to 0.046 (Table~\ref{tab:fit_pars}), satisfying the assumption of a thin disc for successful application of the CF method \citep{2006ApJ...652..518M}. Because the disc luminosity is sufficiently high, the spectral hardening factor $f_{\rm col}$ is well determined, decreasing from 1.645 to 1.527. The best-fit value of $a_{*}$ ranges from 0.71 to 0.86 (Figure~\ref{fig:compare_f} and Table~\ref{tab:fit_pars}), with the average of $0.79\pm0.03$, which is consistent with that obtained from the reflection method \citep[e.g.,][]{2022MNRAS.511.3125J,2023MNRAS.526.6041S}. The derived consistent results of $a_{*}$ indicate that the inner disc has reached the ISCO during the SIMS. The disc density is almost pegged at the upper limit of $10^{22}\,\rm{cm}^{-3}$, indicating a high-density disc reflection in MAXI J1348. It relieves the very high iron abundance required in the previous reflection model, bringing the iron abundance down to $2.5\pm0.8$ times solar abundance, which is consistent with that of \cite{2022MNRAS.513.4869K}. The spectral index $\Gamma$ decreases from $\sim2.4$ to $\sim2.0$, which seems inconsistent with the source becoming softer. Therefore, we attempt to fix the spectral index $\Gamma$ for all observations to its average value and find that this adjustment has a negligible effect on the measurement of other parameters, including the spin. It also has to be noticed that for a few observations, the disc temperature, iron abundance and ionization parameters are fixed at the average value because they are hard to limit (see Table~\ref{tab:fit_pars}), and this does not affect the rest of the parameters. 


We also test the influence of the coronal emissivity parameters on the spectral fitting, especially for the spin measurement. \citet{2014MNRAS.439.2307F} reported that a canonical emissivity index parameter (e.g., the default values $q_1=q_2=3$, $R_{\rm br}=15\,r_{\rm g}$) would lead to the spin being systematically underestimated, particularly when the corona height is greater than 5\,$r_{\rm g}$. A trial of $q_1=0$, $q_2=3$, $R_{\rm br}=5\,r_{\rm g}$ yields a similar fit statistic, with the spin increasing by less than 2\%. To be thorough, another trial of $q_1=8$, $q_2=2$, $R_{\rm br}=5\,r_{\rm g}$ results in a worse fit statistic ($\Delta\chi2 =\sim 100$), and the spin increases by less than 5\%. These results suggest that the emissivity profile has minimal impact on the spin measurement in this source. One of the possible reasons is that the reflection spectrum may not significantly contribute to the constraints on the spin; further details can be found in Section~\ref{dis:err_ana}. 

Considering that the inner disc radius is coupled with the spin, we also test the validity of the assumption that the disc extends to the ISCO, especially given that the spin shows evolution. With the spin of all observations fixed to 0.79, the inferred inner disc radius is found to be unchanged and equal to the ISCO within the margin of error. However, if the spin is set to 0.86, the maximum value measured in this study, the inner disc radius extends beyond the ISCO. This outcome contradicts the fundamental assumption that the disc radius coincides with the ISCO in spin measurements, thereby supporting the reasonableness of a spin value of 0.79.

\begin{figure}
    \includegraphics[width=1.1\columnwidth]{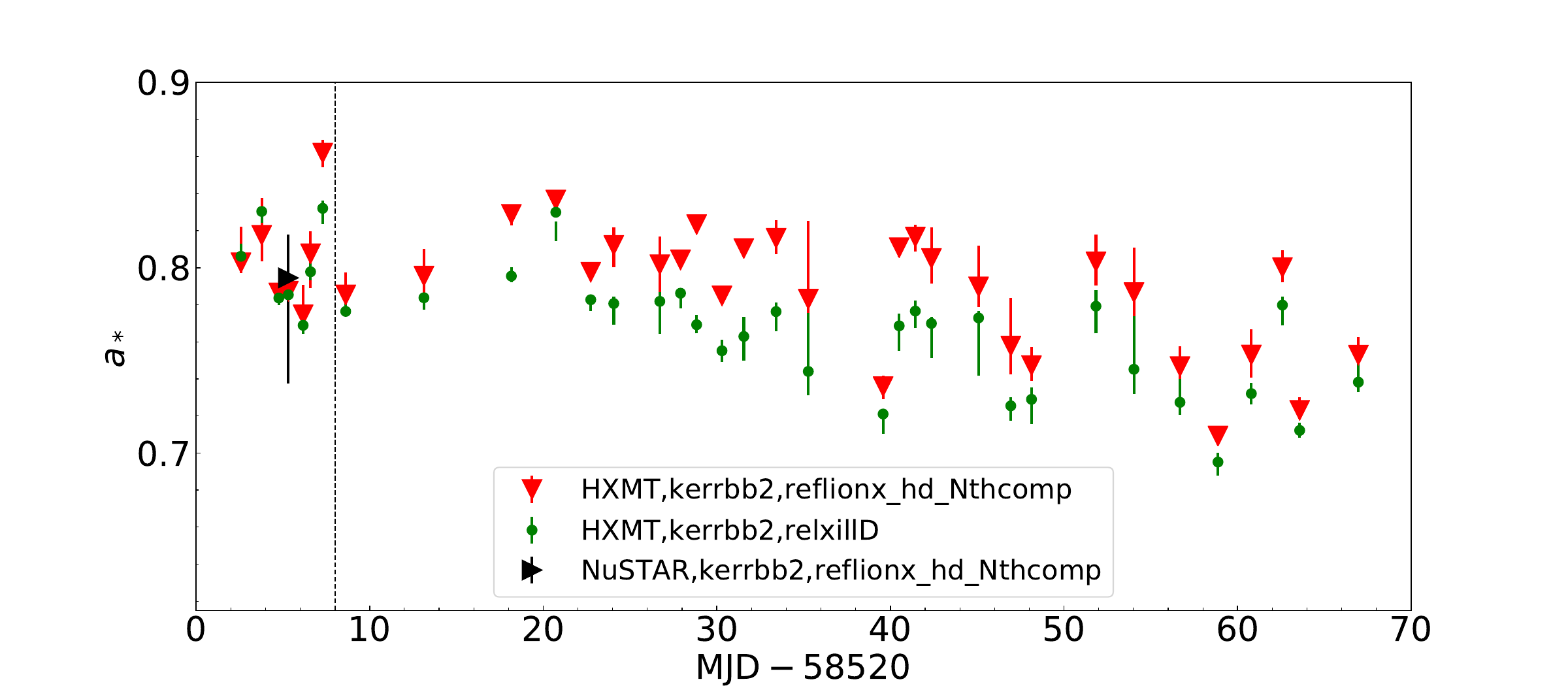}
    \caption{Comparison of the spin derived from different data and models. Red: \textit{Insight}-HXMT spectra fitted with {\tt const*Tbabs*(kerrbb2+NthComp+relconv*reflionx\_hd\_\\Nthcomp)} model, green: \textit{Insight}-HXMT spectra fitted with {\tt const*Tbabs*(kerrbb2+NthComp+relconv*relxillD} model, black: \textit{NuSTAR} spectrum fitted with {\tt const*Tbabs*(kerrbb2+NthComp+relconv*reflionx\_hd\_\\Nthcomp)} model. The vertical line represents the transition from SIMS to SS.}
   \label{fig:compare}
\end{figure}

To investigate the impact of different types of reflection models on spin measurements, we replace the reflection model {\tt reflionx\_hd\_Nthcomp} with the high density relxill family model {\tt relxillD}. Still, we assume canonical emissivity parameters and use the default outer radius for the accretion disc, with the inner disc located at the ISCO. The spin and the spectral index $\Gamma$ of {\tt relxillD} are linked to that of {\tt kerrbb2} and {\tt Nthcomp}, respectively. The reflection fraction, defined as the ratio of the Compton photons emitting towards the disc compared to that escaping to infinity, is restricted to a negative value thereby only the reﬂected component is returned. As shown in Figure~\ref{fig:compare}, the alternative reflection model yields a spin estimation that is consistent with our previous findings, albeit slightly lower. This minor discrepancy could be attributed to the model's disc density, which may still be lower, and the systematically high iron abundance observed during the fitting process, which also contributed to an increased chi-square value. Specifically, the iron abundance is pegged at the upper limit of 10, and the disc density is pegged at the upper limit of  $10^{19}\,\rm{cm}^{-3}$ across all observations, which also supporting the presence of a high-density disc in MAXI J1348.

Moreover, to demonstrate that our spin measurements are not significantly impacted by instrumental effects, we use a simultaneous observation of \textit{NuSTAR} data to validate our results. All the fitting configurations are the same as those used with \textit{Insight}-HXMT data, except that the $\Gamma$ parameter between the two detectors, FPMA and FPMB, is unlinked due to calibration differences \citep{Madsen2015}. A good fit is achieved with no obvious excess in the residuals ($\chi^2/\text{dof} = 1513.5/1896$). As shown in Figure~\ref{fig:compare}, we obtain a consistent spin constraint of $0.79^{+0.02}_{-0.06}$. In addition, the disc density also shows a high value of $1.5^{+2.7}_{-0.9} \times 10^{21}\,\rm{cm}^{-3}$, and the iron abundance is low, at $1.0^{+0.4}_{-0.5}$.


\begin{figure}
    \includegraphics[width=1.1\columnwidth]{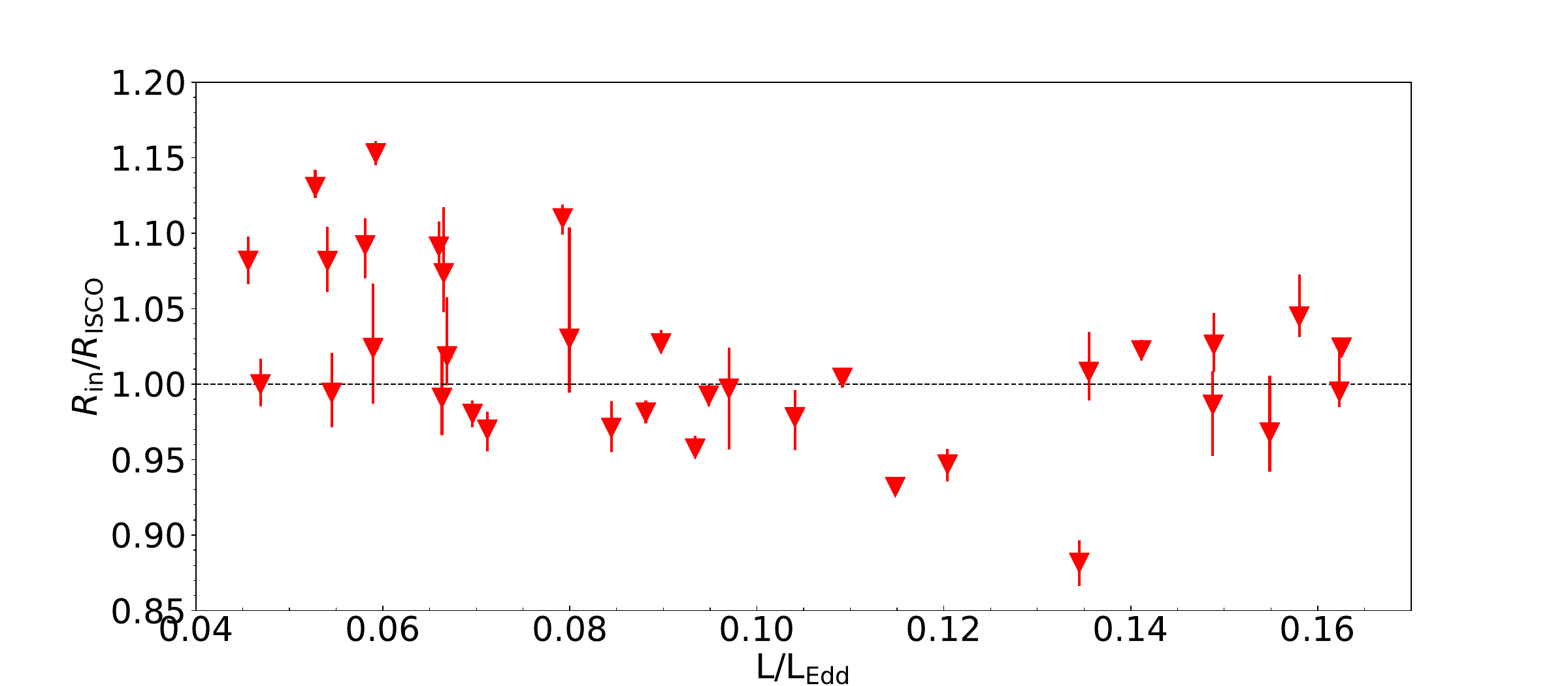}
    \caption{The $R_{\rm{ISCO}}$-scaled inner disc radius, simply derived from the formula of spin vs ISCO radius, as a function of the Eddington-scaled luminosity. $R_{\rm{ISCO}}$ is calculated assuming the spin value to be 0.8, while the Eddington luminosity is computed assuming $M=11\,M_{\odot}$.}
   \label{fig:L_R}
   \end{figure}

\begin{table*}
  \begin{center}
  \renewcommand\arraystretch{1.5}
	\caption{Best-fitting parameters modeled with {\tt const*Tbabs*(kerrbb2+NthComp+relconv*reflionx\_hd\_Nthcomp)}. $a_{*}$ is the spin; $\dot{M}$ is the mass accretion rate in units of $10^{18}~\rm{g~s^{-1}}$; $N_{\rm nthC}$ is the normalization of {\tt NthComp}; $\xi$ is the ionization of the accretion disc; $A_{\rm{Fe}}$ is the iron abundance; $\Gamma$ is the photon index; $kT_{\rm{bb}}$ is the seed photon temperature; $N_{\rm Refl}$ is the normalizaton of {\tt reflionx\_hd\_Nthcomp}; $f_{\rm col}$ is the spectral hardening factor; $l$ is the Eddington-scaled luminosity.}
	\label{tab:fit_pars}
	\resizebox{\textwidth}{!}{
	\begin{tabular}{cccccccccccc}  
	\hline
	ObsID & $a_{*}$ & $\dot{M}$ & $N_{\rm nthC}$ & $\xi$ & $A_{\rm{Fe}}$ & $\Gamma$ & $kT_{\rm{bb}} $ & $N_{\rm Refl}$ & $f_{\rm col}$ & $l$ & $\chi^{2}/v$ \\ 
\hline
 & ($c\rm{J/GM^{2}}$) & ($10^{18}~\rm{g~s^{-1}}$) &  & (erg~cm~s$^{-1}$) & (solar) & & (keV) &  &  &  &  \\
\hline
\startdata
1101$\sim$1103 & $0.803_{-0.006}^{+0.019}$ & $2.03_{-0.16}^{+0.05}$ & $3.3_{-1.0}^{+2.2}$ & $178_{-13}^{+26}$ & $3.5_{-0.9}^{+1.3}$ & $2.361_{-0.009}^{+0.010}$ & $1.1\pm{0.3}$ & $8.7_{-0.8}^{+5.4}$ & 1.645 & 0.162 & 782.6/1159 \\
1501$\sim$1505 & $0.818_{-0.014}^{+0.020}$ & $1.88_{-0.12}^{+0.11}$ & $4.8_{-1.8}^{+1.9}$ & $205_{-14}^{+34}$ & $3.6\pm{0.8}$ & $2.340\pm{0.008}$ & $0.8_{-0.2}^{+0.3}$ & $9.3_{-0.8}^{+2.0}$ & 1.643 & 0.155 & 741.2/1192 \\
1506$\sim$1509 & $0.786_{-0.003}^{+0.004}$ & $2.10_{-0.07}^{+0.03}$ & $1.9_{-0.4}^{+0.9}$ & $190_{-15}^{+35}$ & $2.8_{-0.6}^{+1.0}$ & $2.343_{-0.010}^{+0.012}$ & $1.27_{-0.15}^{+0.02}$ & $7.0_{-0.7}^{+3.2}$ & 1.644 & 0.163 & 749.1/1169 \\
1510$\sim$1513 & $0.787\pm{0.004}$ & $2.06_{-0.07}^{+0.03}$ & $2.0_{-0.4}^{+1.2}$ & $203_{-17}^{+53}$ & $2.5^{*}$ & $2.343_{-0.011}^{+0.010}$ & $1.3_{-0.3}^{+0.2}$ & $7.6_{-0.5}^{+3.5}$ & 1.637 & 0.141 & 787.1/1158 \\
1601$\sim$1603 & $0.775_{-0.008}^{+0.016}$ & $2.09_{-0.13}^{+0.06}$ & $2.0_{-0.3}^{+2.1}$ & $214_{-24}^{+52}$ & $2.5_{-0.5}^{+1.1}$ & $2.339_{-0.016}^{+0.015}$ & $1.0\pm{0.4}$ & $6.7_{-1.0}^{+3.0}$ & 1.639 & 0.158 & 748.0/1090 \\
1604$\sim$1607 & $0.808_{-0.019}^{+0.012}$ & $1.85_{-0.06}^{+0.11}$ & $4.8_{-1.6}^{+1.1}$ & $209_{-21}^{+27}$ & $3.7_{-1.0}^{+1.4}$ & $2.316_{-0.013}^{+0.012}$ & $0.66_{-0.11}^{+0.24}$ & $7.0_{-0.8}^{+1.3}$ & 1.632 & 0.149 & 738.6/1169 \\
1701$\sim$1704 & $0.862_{-0.008}^{+0.007}$ & $1.47\pm{0.03}$ & $8.4_{-0.8}^{+0.9}$ & $244_{-24}^{+33}$ & $2.5^{*}$ & $2.307\pm{0.009}$ & $0.42\pm{0.05}$ & $10.3_{-0.6}^{+1.2}$ & 1.635 & 0.134 & 784.9/1193 \\
1801$\sim$1803 & $0.785_{-0.010}^{+0.012}$ & $1.90\pm{0.06}$ & $3.1_{-0.8}^{+0.9}$ & $202_{-19}^{+31}$ & $2.5^{*}$ & $2.306_{-0.012}^{+0.011}$ & $0.67_{-0.14}^{+0.19}$ & $7.4_{-0.5}^{+0.8}$ & 1.630 & 0.149 & 813.7/1188 \\
2101$\sim$2103 & $0.795_{-0.011}^{+0.015}$ & $1.72_{-0.08}^{+0.07}$ & $2.7_{-0.9}^{+1.2}$ & $174_{-15}^{+12}$ & $2.8_{-0.6}^{+1.0}$ & $2.33\pm{0.02}$ & $0.7_{-0.2}^{+0.3}$ & $8.1_{-0.8}^{+1.4}$ & 1.615 & 0.136 & 812.4/1189 \\
2401 & $0.829_{-0.006}^{+0.005}$ & $1.43_{-0.02}^{+0.04}$ & $4.6_{-0.9}^{+0.6}$ & $197_{-16}^{+17}$ & $4.5_{-1.1}^{+0.5}$ & $2.286_{-0.012}^{+0.013}$ & $0.44_{-0.06}^{+0.11}$ & $7.2_{-0.6}^{+0.8}$ & 1.609 & 0.120 & 761.7/1192 \\
2510$\sim$2512 & $0.837\pm{0.003}$ & $1.322\pm{0.010}$ & $5.52_{-0.12}^{+0.13}$ & $241_{-18}^{+20}$ & $2.5^{*}$ & $2.257_{-0.008}^{+0.009}$ & $0.100_{**}^{+0.011}$ & $9.2_{-0.5}^{+0.7}$ & 1.604 & 0.115 & 747.9/1193 \\
2601$\sim$2602 & $0.798\pm{0.004}$ & $1.381\pm{0.014}$ & $3.36_{-0.11}^{+0.13}$ & $202_{-16}^{+19}$ & $4.1_{-1.1}^{+0.9}$ & $2.199\pm{0.014}$ & $0.10_{**}^{+0.09}$ & $6.4_{-1.1}^{+0.8}$ & 1.588 & 0.109 & 777.2/1125 \\
2701 & $0.812_{-0.012}^{+0.010}$ & $1.28_{-0.04}^{+0.05}$ & $3.7_{-0.8}^{+0.7}$ & $229_{-24}^{+30}$ & $3.5_{-0.9}^{+1.3}$ & $2.28\pm{0.02}$ & $0.443_{-0.09}^{+0.13}$ & $6.7_{-0.7}^{+0.9}$ & 1.587 & 0.104 & 878.0/1082 \\
2801 & $0.80\pm{0.02}$ & $1.22_{-0.05}^{+0.09}$ & $3.1_{-1.3}^{+1.0}$ & $185_{-23}^{+35}$ & $2.5_{-0.4}^{+1.3}$ & $2.29\pm{0.02}$ & $0.52_{-0.14}^{+0.32}$ & $7.0_{-0.5}^{+4.0}$ & 1.577 & 0.097 & 785.0/932 \\
2901 & $0.804\pm{0.004}$ & $1.185_{-0.011}^{+0.012}$ & $4.16_{-0.13}^{+0.20}$ & $218\pm{17}$ & $2.4_{-0.2}^{+0.9}$ & $2.268_{-0.013}^{+0.012}$ & $0.10_{**}^{+0.09}$ & $8.7_{-1.9}^{+0.6}$ & 1.576 & 0.095 & 856.6/1151 \\
3001$\sim$3003 & $0.823_{-0.002}^{+0.005}$ & $1.123_{-0.015}^{+0.010}$ & $4.67_{-0.14}^{+0.19}$ & $253_{-18}^{+17}$ & $2.3_{-0.4}^{+0.2}$ & $2.246_{-0.008}^{+0.013}$ & $0.10_{**}^{+0.05}$ & $9.4_{-1.0}^{+1.1}$ & 1.579 & 0.093 & 797.4/1192 \\
3101$\sim$3104 & $0.785_{-0.004}^{+0.005}$ & $1.163_{-0.015}^{+0.010}$ & $3.12_{-0.15}^{+0.11}$ & $194_{-12}^{+13}$ & $2.3_{-0.5}^{+0.3}$ & $2.25\pm{0.02}$ & $0.18_{-0.08}^{+0.06}$ & $7.2_{-0.5}^{+1.7}$ & 1.571 & 0.090 & 875.0/1178 \\
3201 & $0.810\pm{0.004}$ & $1.088_{-0.013}^{+0.010}$ & $3.90_{-0.13}^{+0.21}$ & $210_{-35}^{+16}$ & $2.3_{-0.3}^{+0.4}$ & $2.259_{-0.013}^{+0.019}$ & $0.10_{**}^{+0.09}$ & $8.4_{-1.3}^{+1.5}$ & 1.571 & 0.088 & 863.6/1122 \\
3601 & $0.816_{-0.009}^{+0.010}$ & $1.03\pm{0.03}$ & $3.1\pm{0.7}$ & $184_{-20}^{+21}$ & $2.3\pm{0.4}$ & $2.34_{-0.02}^{+0.03}$ & $0.41_{-0.10}^{+0.13}$ & $7.7_{-0.5}^{+3.2}$ & 1.569 & 0.084 & 823.6/1052 \\
3801 & $0.78_{-0.02}^{+0.04}$ & $1.04_{-0.11}^{+0.05}$ & $2.8\pm{0.4}$ & $190_{-56}^{+43}$ & $1.8_{-1.0}^{+1.6}$ & $2.24_{-0.05}^{+0.06}$ & $0.20_{-0.10}^{+0.38}$ & $6.4_{-1.6}^{+5.8}$ & 1.563 & 0.080 & 618.7/656 \\
4201 & $0.736_{-0.007}^{+0.006}$ & $1.125_{-0.013}^{+0.016}$ & $1.1_{-0.3}^{+0.2}$ & $213_{-35}^{+50}$ & $2.5^{*}$ & $2.18\pm{0.04}$ & $0.3\pm{0.2}$ & $2.7_{-0.3}^{+0.7}$ & 1.564 & 0.079 & 784.2/1007 \\
4301 & $0.811\pm{0.005}$ & $0.891\pm{0.010}$ & $2.38_{-0.28}^{+0.15}$ & $252_{-28}^{+29}$ & $2.5^{*}$ & $2.21\pm{0.02}$ & $0.22_{-0.06}^{+0.10}$ & $5.5_{-0.4}^{+0.5}$ & 1.555 & 0.070 & 852.8/1076 \\
4401 & $0.817_{-0.008}^{+0.007}$ & $0.869_{-0.013}^{+0.018}$ & $2.3\pm{0.4}$ & $261_{-31}^{+46}$ & $2.5^{*}$ & $2.25\pm{0.02}$ & $0.32_{-0.12}^{+0.11}$ & $5.0\pm{0.4}$ & 1.555 & 0.071 & 835.5/1087 \\
4501 & $0.805_{-0.013}^{+0.017}$ & $0.83_{-0.04}^{+0.03}$ & $1.87_{-0.14}^{+0.29}$ & $193_{-26}^{+32}$ & $0.96_{-0.11}^{+0.68}$ & $2.20_{-0.03}^{+0.02}$ & $0.10_{**}^{+0.05}$ & $9_{-3}^{+2}$ & 1.552 & 0.066 & 830.4/1047 \\
4601 & $0.790_{-0.011}^{+0.022}$ & $0.86_{-0.05}^{+0.03}$ & $1.0\pm{0.4}$ & $218^{*}$ & $1.0_{-0.3}^{+1.7}$ & $2.20_{-0.06}^{+0.09}$ & $0.1_{**}^{+0.3}$ & $7_{-3}^{+7}$ & 1.552 & 0.067 & 519.3/552 \\
4801$\sim$4803 & $0.76_{-0.02}^{+0.03}$ & $0.90_{-0.05}^{+0.03}$ & $0.7\pm{0.2}$ & $218_{-53}^{+77}$ & $0.9_{-0.2}^{+0.5}$ & $2.15_{-0.04}^{+0.06}$ & $0.10_{**}^{+0.09}$ & $5.2_{-1.9}^{+2.4}$ & 1.554 & 0.067 & 760.5/947 \\
4901 & $0.732_{-0.007}^{+0.018}$ & $0.95_{-0.04}^{+0.02}$ & $0.5_{-0.2}^{+0.6}$ & $188_{-31}^{+59}$ & $2.5^{*}$ & $2.22\pm{0.05}$ & $0.262^{*}$ & $2.4_{-0.4}^{+0.6}$ & 1.553 & 0.068 & 692.8/773 \\
5101 & $0.803_{-0.013}^{+0.015}$ & $0.68\pm{0.03}$ & $1.9_{-0.2}^{+0.3}$ & $208_{-30}^{+33}$ & $1.7_{-0.5}^{+1.1}$ & $2.19_{-0.03}^{+0.04}$ & $0.10_{**}^{+0.17}$ & $7_{-2}^{+3}$ & 1.539 & 0.055 & 738.9/724 \\
5501 & $0.79\pm{0.02}$ & $0.76_{-0.05}^{+0.08}$ & $1.7_{-0.2}^{+1.4}$ & $120_{-89}^{+25}$ & $2.8_{-1.2}^{+2.2}$ & $2.19_{-0.05}^{+0.06}$ & $0.20_{-0.10}^{+0.67}$ & $5.0_{-1.1}^{+1.9}$ & 1.545 & 0.059 & 465.1/527 \\
5601$\sim$5604 & $0.747_{-0.013}^{+0.011}$ & $0.81_{-0.02}^{+0.03}$ & $0.7_{-0.4}^{+0.3}$ & $138_{-18}^{+20}$ & $2.5^{*}$ & $2.17\pm{0.04}$ & $0.4_{-0.2}^{+0.5}$ & $2.8_{-0.5}^{+0.3}$ & 1.546 & 0.058 & 780.9/870 \\
5701$\sim$5703 & $0.711_{-0.007}^{+0.004}$ & $0.877\pm{0.009}$ & $0.113_{-0.016}^{+0.057}$ & $59_{-27}^{+65}$ & $2.5^{*}$ & $2.16\pm{0.09}$ & $0.262^{*}$ & $2.4_{-1.0}^{+1.5}$ & 1.543 & 0.060 & 673.6/799 \\
5801$\sim$5802 & $0.753_{-0.012}^{+0.014}$ & $0.741_{-0.021}^{+0.011}$ & $0.84_{-0.14}^{+0.08}$ & $155\pm{17}$ & $2.2_{-0.7}^{+0.4}$ & $2.09_{-0.03}^{+0.04}$ & $0.20_{-0.10}^{+0.15}$ & $2.5_{-0.3}^{+1.2}$ & 1.541 & 0.054 & 800.3/974 \\
5901$\sim$5904 & $0.800_{-0.008}^{+0.009}$ & $0.591_{-0.017}^{+0.012}$ & $0.70\pm{0.06}$ & $188_{-29}^{+30}$ & $1.1\pm{0.2}$ & $2.07\pm{0.03}$ & $0.10_{**}^{+0.10}$ & $3.0_{-0.6}^{+1.0}$ & 1.529 & 0.047 & 858.6/1055 \\
6001$\sim$6005 & $0.723_{-0.004}^{+0.007}$ & $0.758_{-0.014}^{+0.008}$ & $0.36_{-0.05}^{+0.10}$ & $152_{-18}^{+23}$ & $2.3_{-0.8}^{+0.4}$ & $1.99_{-0.04}^{+0.06}$ & $0.4_{-0.3}^{+0.2}$ & $1.5_{-0.2}^{+1.1}$ & 1.535 & 0.053 & 875.1/985 \\
6101 & $0.753_{-0.009}^{+0.010}$ & $0.647_{-0.012}^{+0.013}$ & $0.77_{-0.21}^{+0.11}$ & $162_{-18}^{+21}$ & $2.5^{*}$ & $2.14_{-0.04}^{+0.05}$ & $0.25_{-0.10}^{+0.21}$ & $2.6_{-0.3}^{+0.4}$ & 1.527 & 0.046 & 733.8/883 \\
\hline
	\end{tabular}
	}
	\end{center}
\footnotesize{$^{*}$ For some observations, as $A_{\rm{Fe}}$ and $kT_{\rm bb}$ are} difficult to be constrained, we fix it at the average value. The error of $\xi$ is very large in P021400204601, thus we fix it at the best fitting value. See Section~\ref{sec:spectral properties} for more details.\newline
\footnotesize{$^{**}$ The error of the parameter is pegged at the upper or lower limit.}\newline 
\end{table*}

\subsection{Error analysis}

The dominate error budge in our determination of the spin is the combined uncertainties of the three key parameters $M$, $i$ and $D$ . Following the prescription described in \citet{2011ApJ...742...85G} and \citet{2021MNRAS.504.2168G}, we perform Monte Carlo (MC) simulations to analyze the error. Since the mass function is unknown, $M$, $i$ and $D$ are assumed independent and normally distributed (i.e. $M = 11 \pm 2\,M_{\odot}$, $i=29.3\pm3.2\,^{\circ}$, and $D = 3.39 \pm 0.34$ kpc. Here we adopt a larger error to ensure coverage of the asymmetric error range, especially when the upper and lower error limits are asymmetric). We sample 6000 triplets of $M$, $i$ and $D$, which are distributed in a uniform grid throughout the allowed parameter space. For each point in the grid, a look-up table is calculated. Then a representative spectrum (ObsID P021400203601, since its derived $a_*$ is near the average of all the observations) is re-fitted with the above combined model to determine $a_*$. Folding all of the runs together, a probability distribution of $a_*$ based on the selected spectrum is obtained. The errors in $M$, $i$ and $D$ account for $\Delta a_* \approx0.13$ at 68 percent confidence. Using the average value of all the observations and also considering the systematic error from all of the observations, we constrain $a_*$ to be $0.79\pm0.13$ at the 1$\sigma$ level of confidence,  confirming a moderately high spin.

\section{Discussion}

\label{sec:dis}

In this paper, we report the results of a broad-band spectral analysis of MAXI J1348 during its SIMS and SS in the 2019 outburst based on the \textit{Insight}-HXMT data. We use a joint-fitting method for the continuum and reflection components to determine the spin of MAXI J1348-630. We also investigate the impact of other factors on the spin measurements.

\subsection{Spin measurement}
\label{dis:spin_meas}

We used joint fitting of continuum and reflection components, with a total of 31 \textit{Insight}-HXMT observations that spanned 65 days and covered the broadband energy band of 1--150 keV. From this analysis, we estimated the average spin of MAXI J1348 to be $0.79 \pm 0.13$ with a 1$\sigma$ confidence level. Key input parameters and their uncertainties are as follows: the mass $M = 11 \pm 2\,M_{\odot}$, the inclination $i=29.3\pm3.2\,^{\circ}$, and the distance $D = 3.39 \pm 0.34$ kpc. The spin measurement is further validated through simultaneous observation with \textit{NuSTAR} data and the application of an alternative reflection model, demonstrating that our spin results are independent of both the chosen model and the dataset utilized.


Thus far, there have been six reports on the spin measurement of MAXI J1348, with most yielding moderately high spin values that align with our result of $0.79 \pm 0.13$. We listed them in Table~\ref{tab:spins}. \citet{2022MNRAS.511.3125J} used the reflection model \textit{relxillCp} and combined the four \textit{NuSTAR} observations to measure the spin as $0.78 \pm 0.04$. These observations covered the IMS and SS, during which the inner radius of the disc was stable at the ISCO. Similarly, \citet{2023MNRAS.526.6041S} used the relativistic reflection model {\tt relxill} with six \textit{Insight}-HXMT observations (in the energy band of 1 to 50 keV) in the IMS and SS, determining the spin of MAXI J1348 to be $0.82^{+0.04}_{-0.03}$. \citet{2022MNRAS.513.4869K} utilized a continuum and reflection components joint fitting method with the \textit{NICER} and \textit{NuSTAR} data during the SS, covering the 0.6--79\,keV energy band. They constrained the spin to be $0.80\pm0.02$, with a lighter black hole mass of $8.7\pm 0.3\,M_{\odot}$, a larger inclination of $36.5^{\circ} \pm 1.0^{\circ}$, and a shorter distance of 2.2\,kpc. All of these results are consistent with our report on the moderately high spin. 

{\citet{2024ApJ...969...40D} also applied the reflection method based on the relxill family model to estimate the spin of MAXI J1348 from all \textit{NuSTAR} observations covering the HS, IMS and SS. They found, however, an extreme black hole spin of $0.977^{+0.017}_{-0.055}$ and a high inclination of $52^{+8}_{-11}[^{\circ}]$ from the combined posterior probability distributions. They attributed the higher derived spin value, which was greater than those measured by \citet{2022MNRAS.511.3125J}, to the free variation of the emissivity parameters. However, some observations still yield low spin and inclination values, regardless of whether the Gaussian features are included in the fitting process. Therefore, we suspect that the spectrum fitting, particularly of the accuracy of the continuum, may influence the reflection component and the spin measurement, which may explain the inconsistency of the spin measurement with our results (for details, see Section~\ref{dis:err_ana}). \citet{2023RAA....23a5015M} also derived an extreme black hole spin of $>0.97$ using three \textit{AstroSat} observations covering the energy band of 0.8--25\,keV with a joint-fitting method. We suspect that this discrepancy may be due to the relatively narrow energy band ($<25$\,keV) obtained from the \textit{AstroSat} observations which would make the fitting more easily affected by the systematic biases associated with the telescope.

The most deviant result originates from \citet{2023MNRAS.522.4323W}, who employed the CF method using five observations from \textit{Insight}-HXMT within the 2–20 keV energy band. They estimated the spin of MAXI J1348 to be as low as $0.41^{+0.13}_{-0.50}$, although they used the same system parameters ($M$, $i$, and $D$) as in our work. The discrepancy may be attributed to the fixed $f_{\rm col}=1.6$ during their spectral fitting with the {\tt kerrbb} model. Moreover, given that the characteristic temperature of the disc is around 1\,keV \citep[e.g.,][]{2006ApJ...652..518M}, utilizing data from the 2--20 keV energy band might not accurately constrain the emission from the disc. This limitation could also affect the spin measurement when employing the CF method. Furthermore, when they examined a range of $f_{\rm col}$ (from 1.5 to 1.7) to test its impact on the spin measurement, the spin consistently inclined towards a lower value. It is noteworthy that the parameters—accretion rate \( \dot{M} \), spin \( a_* \), and spectral hardening factor \( f_{\rm col} \)—are coupled in the {\tt kerrbb} model \citep{2005ApJS..157..335L}, and the simultaneous fitting of these parameters may not yield accurate measurement results. In our work, we adopt the {\tt kerrbb2} model to avoid this problem, see Section~\ref{sec:spectral properties}. Moreover, in the following section (Section~\ref{dis:err_ana}), we will demonstrate that fixing $f_{\rm col}$ at a constant value would lead to a distortion of the spin measurement.

\begin{table*}
\small
\caption{Comparative of spin measurements in different reports.}
\resizebox{\textwidth}{!}{
\label{tab:spins}
\centering
\begin{tabular}{ccccccc}
\toprule
spin value & satellite & states & method & high disc density? & $f_{\rm col}$ & reference \\
\hline
$0.78 \pm 0.04$ & \textit{NuSTAR} & IMS and SS & reflection fitting & Yes & - & \citet{2022MNRAS.511.3125J} \\
$0.80 \pm 0.02$ & \textit{NICER} and \textit{NuSTAR} & SS & reflection and continuum joint fitting & No & 1.7 & \citet{2022MNRAS.513.4869K} \\
$0.82^{+0.04}_{-0.03}$ & \textit{Insight}-HXMT & IMS and SS & reflection fitting & Yes & - & \citet{2023MNRAS.526.6041S} \\
$>0.97$ & \textit{AstroSat} & SS & reflection and continuum joint fitting & No & 1.7 & \citet{2023RAA....23a5015M} \\
$0.41^{+0.13}_{-0.50}$ & \textit{Insight}-HXMT & SS & continuum fitting & No & 1.6 & \citet{2023MNRAS.522.4323W} \\
$0.977^{+0.017}_{-0.055}$ & \textit{NuSTAR} & HS, IMS and SS & reflection fitting & No & - & \citet{2024ApJ...969...40D} \\
$0.79 \pm 0.13$ & \textit{Insight}-HXMT & IMS and SS & reflection and continuum joint fitting & Yes & 1.53--1.65 & This work \\
\hline
\end{tabular}}
\end{table*}


\subsection{Systematic error analysis of spin measurement}
\label{dis:err_ana}

The biggest obstacle left to estimate black hole spin via the CF method would be the uncertainty due to the assumed hardening factor of the disc spectrum. Non-blackbody effects are usually modeled in terms of the spectral hardening factor, which is found to be inversely correlated with the spin and could not be fitted with the spin and the mass accretion rate at one time from {\tt kerrbb} \citep{2005ApJS..157..335L}. Independent estimation of the hardening factor from more reliable spectral models of disc atmospheres (e.g. {\tt BHSPEC}) is recommended \citep{2006ApJ...652..518M}. Indeed, when we replace {\tt kerrbb2} with the variable $f_{\rm col}$ by {\tt kerrbb} with $f_{\rm col}$ fixed at 1.6, the spin shows an obvious declining trend across time (Figure~\ref{fig:compare_f}). Since the physical spin evolution for stellar mass BHs is on a timescale of $10^9$\,years \citep{2021ApJ...906..105C}, it is reasonable to believe that inaccuracies in the hardening factor may result in biased spin estimates. Indeed it has been suggested that $f_{\rm col}$ is relatively strongly influenced by the accretion rate \citep{2000MNRAS.313..193M, 2006ApJ...647..525D}. Hence, a correct value of $f_{\rm col}$ is crucial for measuring the spin using the CF method.

Thanks to the multiple \textit{Insight}-HXMT observations of MAXI J1348, we can measure the spin value of this source numerous times, yielding a more accurate result and enabling the analysis of physical factors that could influence the measurement. As illustrated in Figure~\ref{fig:compare_f}, we obtained a more accurate spin measurement with {\tt kerrbb2} compared with that from {\tt kerrbb}. However, to be noticed, even when considering the effect of $f_{\rm col}$, the spin exhibits slight variability, suggesting that some underlying factors might have not been fully taken into consideration. The most straightforward one might be the suspicion of the precision of $f_{\rm col}$ in {\tt kerrbb2}, which could be influenced by factors such as the assumption of $\alpha$-viscosity prescription and the sampling granularity of the look-up tables for $f_{\rm col}$. To quantify the magnitude of the systematics induced by the uncertainty in $f_{\rm col}$, we replace {\tt kerrbb2} with {\tt kerrbb} with the spin fixed to 0.8 to derive the best-fit $f_{\rm col}$. As shown by Figure~\ref{fig:compare_f}, the best-fit $f_{\rm col}$ decreases more steeply compared with that derived from {\tt kerrbb2}. Given their similar fit statistics, we cannot rule out the possibility that the decreasing trend in spin observed with the {\tt kerrbb2} model may result from an imprecise estimate of $f_{\rm col}$. However, one thing is certain: a more accurate evolution of $f_{\rm col}$ would eliminate the observed decrease in spin, demonstrating that the hardening factor is strongly inversely correlated with spin. 

In addition to the impact of $f_{\rm col}$ on spin measurement, other potential physical factors may also need to be taken into account. For example, the corona structure is more complicated than the current model; multiple coronae have been reported in some cases (e.g., Swift J1727.8--1613, \citealt[][]{Peng2024}; MAXI J1820+070, \citealt[][]{Ma2023, Lucchini2023}). Actually, in MAXI J1348, \citet{2021MNRAS.501.3173G} and \citet{2022ApJ...927..210Z} reported that there are two Comptonization components shown in SIMS, the single-coronal modeling may lead to an artificial broadening of the iron line distortion, which would lead to overestimated the spin. Moreover, the reflection component from the corona-illuminated disc can be scattered again by the corona, leading to further distortion and a broader iron line \citep{Steiner2017}. Additionally, other factors influencing the disc structure can also introduce errors in spin measurements \citep[e.g.,][]{Riaz2020, Zdziarski2024a, Zdziarski2024b}. Therefore, future work is needed to address uncertainties in spin measurements by focusing on these issues, which are beyond the scope of this paper.

Besides the physical factors discussed above, we also explore whether the changing spin can be attributed to variations in the inner disc radius, especially given the long-standing debate surrounding the evolution of the inner disc radius. To investigate this, we calculate the inner disc radius using the spin vs ISCO radius formula \citep[][]{1997ApJ...482L.155Z} for each measured spin. We then scale the radius assuming a spin of 0.8, the reported spin value, and plot it as a function of the Eddington-scaled luminosity. As shown in Figure~\ref{fig:L_R}, the theoretically calculated inner disc radius approaches the ISCO for most luminosities, with only a slight increase observed at low luminosities around $\sim5\%$\,${L}_{\rm{Edd}}$. The position of the inner disc radius relative to the ISCO, particularly in the luminosity range of $\sim0.1\%$--10$\%$\,${L}_{\rm{Edd}}$ for the HS, has been a subject of considerable debate \citep[e.g.,][]{2015ApJ...813...84G, 2021ARA&A..59..117R}. Our results indicate that, the inner disc during the late SS phase may also experience a slight truncation, which could lead to a decrease in spin.


When employing the method of jointly fitting the continuum and reflection components to measure spin, it is essential to consider which component predominantly influences the spin measurement. This is especially relevant in \textit{Insight}-HXMT data, where the spectral bins near the iron line are fewer than those of the continuum. To explore this effect further, we conducted a test by unlinking the spin between {\tt kerrbb2} and {\tt reflionx\_hd\_Nthcomp}. We found that the spin derived from {\tt kerrbb2} remained nearly unchanged, while the spin obtained from {\tt reflionx\_hd\_Nthcomp} was consistent but exhibited larger uncertainties. Additionally, we fixed the spin of {\tt reflionx\_hd\_Nthcomp} at 0.98, allowing the spin of the continuum to vary. The resulting spin values were similar, with comparable fit statistics. Based on these findings, we suggest that the fit is predominantly driven by the continuum component, with the reflection component also contributing to the determination of spin. Caution is advised when relying solely on the relativistic reflection method to constrain the spin from spectra that exhibit a strong (not too weak) continuum component. It is crucial to ensure that artificial features introduced by instrumental effects do not compromise the continuum fitting, as this could distort the shape of the reflection component and, consequently, impact the accuracy of the spin measurement.

\subsection{High disc density}
\label{dis:disc_density}

We also find significant evidence of high-density disc reflection in MAXI J1348 with the {\tt NthComp} continuum-based reflection model {\tt reflionx\_hd\_Nthcomp}. The disc density almost reaches $10^{22}\,\rm{cm}^{-3}$, which brings the iron abundance down to $\sim2.5$ times solar value, relieving the super-solar abundance required in other reflection models with a constant density up to $10^{19}\,\rm{cm}^{-3}$ (e.g. {\tt relxillD}). A high-density model allows $A_{\rm Fe}$ to drop because it accounts for the additional effect that the high disc density produces extra thermal emission by the increase in free-free absorption and heating of the disc \citep[e.g.,][]{Garcia2016,2018ApJ...855....3T}.  Our derived iron abundance is consistent with that reported by \cite{2021MNRAS.508..475C}, which uses the same high-density reflection model, and \cite{2022MNRAS.513.4869K}, which constrains $A_{\rm Fe}$ from reflection component in the hard state spectra with less influence from the soft thermal emission. Thus, it is valid to conclude that the iron abundance of MAXI J1348 is larger, but comparable to the solar value. In addition, the higher disc density also impacts other physical parameters, such as the disc inclination and the inner radius of the disc \cite[e.g.,][]{2018ApJ...855....3T, Jiang2019b}, which may affect the spin measurement.

\section{Conclusions}
\label{sec:con}

We analyzed the 1--150 keV \textit{Insight}-HXMT data from a total of 31 observations during the SIMS and SS of MAXI J1348. We use the joint-fitting method of the continuum and reflection components in the energy band of 1 to 150 keV to constrain the spin of this source, and we analyze possible systematic errors in the spin measurements. Our main findings are:

1) By giving the $M=11\pm2\,M_{\odot}$, $i=29.3\pm3.2\,^{\circ}$, and $D=3.39\pm0.34$\,kpc, the average spin of MAXI J1348 is constrained as $0.79\pm0.13$, which is consistent with most of the previous results.

2) We analyzed potential systematic errors in spin measurements. The varying hardening factor can partially account for the measured spin variation during the outburst. Additionally, the elucidation of spin's slight evolution throughout the outburst may be attributed to the accuracy of the hardening factor and the effects of other physical factors, including the complex structures of the corona and their physical interaction process with the reflection photons. 

3) Our results suggest that the selection of different models may have a minor impact on spin measurements, potentially due to constraints imposed by other parameters within the models. However, selecting different datasets does not appear to significantly influence the spin measurement.

4) The accretion disc density of MAXI J1348 is higher, approximately $10^{22}\,\rm{cm}^{-3}$, which can effectively reduce the iron abundance to approximately 2.5 times the solar value.

\acknowledgments
We thank the anonymous referee for useful comments that helped us improve the paper. We thank Michael Parker for providing the reflection model. We also thank Roberto Soria for the useful discussions. We thank Mariano M\'endez and Yuoli Tuo for their help with the MCMC analysis. This work made use of data from the \textit{Insight}-HXMT mission, a project funded by China National Space Administration (CNSA) and the Chinese Academy of Sciences (CAS). This work is supported by the National Key R\&D Program of China (2021YFA0718500). We acknowledges funding support from the National Natural Science Foundation of China (NSFC) under grant Nos. U2038102, 12122306, U1838108, U1838115, U1838202, U1938104, U2031205, U2038103, 12333007, 11733009 and 12027803 the CAS Pioneer Hundred Talent Program Y8291130K2, the Scientific and technological innovation project of IHEP Y7515570U1 and the International Partnership Program of Chinese Academy of Sciences (Grant No.113111KYSB20190020). RM acknowledges support from the China Postdoctoral Science Foundation (Grant No. 2024M751755).

{\it Facilities:} \textit{Insight}-HXMT

\clearpage

\end{document}